\documentstyle[aps,psfig,12pt]{revtex}
\def\beq{\begin{equation}}
\def\eeq{\end{equation}}

\def\bea{\arraycolsep .1em \begin{eqnarray}}
\def\eea{\end{eqnarray}}
\def\Tr{{\rm Tr}}
\def\step{\vspace{.5em}}
\def\bigstep{\vspace{1.5em}}

\def\X{{X}}

\def\eq#1{(\ref{#1})}

\def\Eq#1{Eq.~(\ref{#1})}

\def\s0#1#2{\mbox{\small{$ \frac{#1}{#2} $}}}
\def\0#1#2{\frac{#1}{#2}}

\makeatletter

\renewenvironment{thebibliography}[1]
         {{\bf \noindent References}\\[-1ex]\frenchspacing\small
          \begin{list}{[\arabic{enumi}]}
         {\usecounter{enumi}\parsep=2pt\topsep 0pt
         \settowidth{\labelwidth}{[#1]}
         \leftmargin=\labelwidth\advance\leftmargin\labelsep
         \rightmargin=0pt\itemsep=0pt\sloppy}}{\end{list}}

\makeatother

\begin{document}
\begin{center}

\thispagestyle{empty}

{\normalsize\begin{flushright}
CERN-TH-2002-213\\
FAU-TP3-02-24\\[10ex] \end{flushright}
}

\mbox{\large \bf Wilsonian flows and background fields}
\\[6ex]

{Daniel F. Litim
\footnote{Daniel.Litim@cern.ch}
and 
Jan M.~Pawlowski 
\footnote{jmp@theorie3.physik.uni-erlangen.de}}
\\[4ex]
{${}^*${\it 
Theory Division, 
CERN\\
CH-1211 Geneva 23.
}\\[2ex]${}^\dagger${\it 
Inst. f\"ur Theoretische Physik III, Universit\"at Erlangen,\\ 
Staudtstra\ss e 7, D-91054 Erlangen, Germany.}}
\\[10ex]

{\small \bf Abstract}\\[2ex]
\begin{minipage}{14cm}{\small 
    We study exact renormalisation group flows for background field
    dependent regularisations. It is shown that proper-time flows are
    approximations to exact background field flows for a specific
    class of regulators. We clarify the r$\hat{\rm o}$le of the
    implicit scale dependence introduced by the background field. Its
    impact on the flow is evaluated numerically for scalar theories at
    criticality for different approximations and regularisations.
    Implications for gauge theories are discussed.}
\end{minipage}
\end{center}

\newpage
\pagestyle{plain}
\setcounter{page}{1}
\noindent 
{\bf 1. Introduction}\step

Renormalisation group (RG) techniques are pivotal for the study of
theories which are strongly coupled and/or have diverging correlation
length. For practical purpose renormalisation group flows should be
easily accessible by analytical means and should have good numerical
stability within given truncations.  Such flows provide a promising
starting point for the study of more involved theories like
(non-)Abelian gauge theories or even quantum gravity.  The formulation
of gauge theories is simplified by using background fields. Within the
ERG approach background field techniques have been applied to
non-Abelian gauge theories
\cite{Reuter:1994kw,Reuter:1997gx,Reuter:1995tr,Reuter:1996be,Pawlowski:1998ch,Litim:1998qi,Falkenberg:1998bg,Litim:1998nf,Freire:1996db,Freire:2000bq,Pawlowski:2001df,Bonini:2001xm,Gies:2002af,Litim:2002ce},
to the superconducting phase transition
\cite{Reuter:1994sg,ScalarQED-N} and to a study of UV fixed points of
Euclidean quantum gravity \cite{QuantumGravity}.\step

The full power of the background field method can be exploited by
identifying the background field with the mean field. Then, background
field flows offer remarkable simplifications for both analytical and
numerical implementations. Analytic methods like heat kernel
techniques are applicable and simplify the evaluation of the
operator traces
\cite{Reuter:1994kw,Reuter:1997gx,Pawlowski:2001df,Gies:2002af,Litim:2002ce}.
However, identifying the background field with the mean field entails
an inherent approximation on the flow: the outcome of an integration step
provides an approximation to the input for the next integration step.
Proper-time flows \cite{Liao:1997nm} are subject to a similar
approximation \cite{Litim:2001hk,Litim:2001ky,Litim:2002xm}, the main
difference being that an additional contribution to the flow due to
the field dependence of the regularisation is also neglected
\cite{Litim:2002xm}. For these reasons, it is important to investigate
the implications and limits of these approximations, both for ERG
flows with background fields, and related generalised proper time
flows.  \step

In the present work we address the details of such an approach. We
sketch the derivation of a general background field flow in scalar
theories, which serve as a testing ground for the properties and
limits of the background field approach.  We detail their connection
to proper time flows derived in \cite{Litim:2002xm} and discuss the
inherent approximation from a formal point of view. These
approximations are linked to terms proportional to the scale
derivative of the full inverse propagator.  Their impact is studied
numerically by a calculation of critical exponents in $3d$ scalar
theories to leading order in a derivative expansion. Our results for
different regularisations and approximations are discussed in the
light of their conceptual differences. Implications for gauge theories
are emphasised at various places.  \bigstep

\noindent
{\bf 2. Generalities}\step 

We briefly summarise the key steps of the derivation of the flow
equation within the background field formulation. Consider the
generating functional of a scalar theory with classical action
$S[\phi]$ and a cut-off term $\Delta S_k[\phi,\bar\phi]=\s012 \int
(\phi-\bar\phi) R\ (\phi-\bar\phi)$. Typically, the infra-red
regulator $R$ tends to a mass for small momenta $p^2/k^2\ll 1$ and
decays exponentially for large momenta $p^2/k^2\gg 1$ . In $\Delta
S_k$, $\phi$ is the full field and the background field $\bar\phi$ is
an auxiliary field at our disposal.  A common choice for the
background field $\bar\phi$ is the mean field:
$\bar\phi=\langle\phi\rangle_J$. Note, that in gauge theories
formulated in background field dependent gauges even the gauge fixed
action $S$ depends on both, $\phi$ and $\bar\phi$. The full field
$\phi=\bar\phi+(\phi-\bar\phi)$ is split into the background field
$\bar\phi$ and the fluctuation field $(\phi-\bar\phi)$. In general,
only these fields transform homogeneously under renormalisation group
transformations. We allow for background field dependent regulator $R$
and define the regularised generating functional
\begin{eqnarray}
Z_k[J,\bar\phi]=
\int d\phi 
\exp\left(-S[\phi]-\Delta S_k[\phi,\bar\phi]+\int J(\phi-\bar\phi)\right)\,.
\label{gen}\end{eqnarray}
The propagating field is the fluctuation field $(\phi-\bar\phi)$, which
is coupled to the external current.  The flow equation for the
generating functional in \eq{gen} is given by $\partial_t
Z[J,\bar\phi]=-\langle \partial_t \Delta S_k[\phi,\bar\phi]
\rangle_J$. This equation depends linear on the propagator of the
fluctuation field $(\phi-\bar\phi)$, as necessary for one loop exact flows 
\cite{Litim:2002xm}. The flow of $Z_k$ can be turned
into a flow for the effective action $\Gamma_k[\phi,\bar\phi]=\int
J(\phi-\bar\phi)-\ln Z[J;\bar\phi]+ \Delta S_k[\phi,\bar\phi]$:
\begin{eqnarray}
\partial_t \Gamma_k[\phi,\bar\phi]=
\s012 \Tr\, {1\over \Gamma^{(2,0)}[\phi,\bar\phi]+R}\partial_t R, 
\label{flow}\end{eqnarray}
where 
$$
\Gamma^{(n,m)}[\phi,\bar\phi]={\delta^n\over (\delta\phi)^n}
{\delta^m\over (\delta\bar\phi)^m }\Gamma_k[\phi,\bar\phi]\,.
$$
The flow \eq{flow} is both consistent and complete, in the sense
coined in \cite{Litim:2002xm}. It is a straightforward task to show
that standard perturbation theory is contained in the integrated flow
\eq{flow}. For these considerations, the background field acts as a
spectator. The next question is how the effective action
depends on $\phi$ and $\bar\phi$. If $R$ is independent of $\bar\phi$,
the effective action only depends on the full field $\phi$. In case we
take a $\bar\phi$-dependent $R$, the effective action is {\it not} a
functional of the full field $\phi$, but a functional of both fields.
This can be summarised in the equation
\begin{eqnarray}
\Gamma^{(0,1)}_k[\phi,\bar\phi]=\s012 \Tr 
{1\over \Gamma^{(2,0)}[\phi,\bar\phi]+R}
\0{\delta R}{\delta\bar\phi }\,, 
\label{phi_*}\end{eqnarray}
which controls the background field dependence of the effective
action. \Eq{phi_*} resembles the flow equation \eq{flow}.  The term on
the right-hand side vanishes for $R\to 0$. In gauge theories the
situation is slightly more complicated. As mentioned before, the
action $S$ depends on both, the fluctuation field and the background
field, in background field dependent gauges. Consequently \eq{phi_*}
receives additional contributions from $ \langle
S^{(0,1)}(\phi,\bar\phi)\rangle \neq 0$. These terms are even present
for vanishing regulator $R=0$, 
e.g.~\cite{Litim:1998qi,Litim:1998nf,Freire:2000bq}. An exception 
is provided by axial gauges, where the background field only
enters the regulator term \cite{Litim:2002ce}.\step

Now we want to find a particularly simple form of the flow \eq{flow},
subject to an appropriate choice of the class of regulators $R$. The
flow \eq{flow} depends on the two operators $R$ and
$\Gamma_k^{(2,0)}[\phi,\bar\phi]$. For all purposes, in particular for
its numerical treatment, we would like to choose a regulator $R$ which
commutes with $\Gamma^{(2,0)}[\phi,\bar\phi]$, or at least comes as
close as it gets.  The obvious choice would be to take $R$ as a
function of $\Gamma_k^{(2,0)}[\phi,\bar\phi]$. However, the dependence
on $\phi$ implies that \eq{flow} would then contain additional flow
terms to any loop order, as $\Gamma_k^{(2,0)}[\phi,\bar\phi]$ depends
on arbitrary powers of $\phi$. This spoils the applicability of the
flow. To maintain \eq{flow}, the regulator $R$ only can depend on
$\Gamma_k^{(2)}[\bar\phi,\bar\phi]$. There are other choices, guided
by renormalisation group arguments, which lead to regulators that
depend only linearly on $\Gamma_k^{(2,0)}[\bar\phi,\bar\phi]$, see
\cite{Pawlowski:2001df}. We parametrise the regulators as
\beq\label{R-background}
R(\bar x)= \bar x\, r(\bar x)
\eeq 
and define 
\begin{eqnarray}\label{defofx}
x:=\Gamma_k^{(2,0)}[\phi,\phi]
\end{eqnarray} 
and $\bar x=x[\phi=\bar\phi]$. The corresponding flow is
\begin{eqnarray}
\partial_t \Gamma_k[\phi,\bar\phi]=
\s012 \Tr\, {1\over \Gamma^{(2,0)}[\phi,\bar\phi]+
R[\bar x]}\, \left(\partial_{\bar x} 
R[\bar x]\,\partial_t \bar x-2 \bar x^2\, \partial_{\bar x} r[\bar x]\right)
\label{fullbackflow}\end{eqnarray}
\Eq{fullbackflow} can be used in two ways. First we can expand it
about a background field $\bar\phi$ which has physical importance,
e.g.\ the vacuum configuration. Such a procedure should stabilise the
flow as already the chosen expansion points contains some non-trivial
physical information. Of course this means that this information has
to be at hand. Second, we can use \eq{fullbackflow} at
$\bar\phi=\phi$, where the background field is the mean field.  In
this case we deal with the flow of the effective action
\begin{eqnarray} 
\Gamma_k[\phi]:=\Gamma_k[\phi,\phi]. 
\label{onefield}\end{eqnarray} 
Then, the flow on the right-hand side of \eq{fullbackflow} solely
depends on the propagator $\Gamma_k^{(2,0)}[\phi,\phi]$ and its
$t$-derivative. Inserting this in \eq{fullbackflow} at $\bar\phi=\phi$
we arrive at
\begin{eqnarray}\label{flowgeneral}
\partial_t \Gamma_k[\phi] = 
- \Tr\,{x\, r' \over 1+r }
+\s012 \, \Tr\, {r+ x\,r' \over x(1+r) }\,\dot x\,,  
\end{eqnarray}
where $\dot{x}=\partial_t x$ and $r'=\partial_x r$.  We rush to add
that \eq{flowgeneral} is not closed. The propagator
$\Gamma^{(2,0)}[\phi,\phi]$ is that of the fluctuation field, which is
not identical to
$$\Gamma^{(2)}_k[\phi]=\Gamma^{(2,0)}[\phi,\phi]+ 2
\Gamma^{(1,1)}[\phi,\phi]+\Gamma^{(0,2)}[\phi,\phi]$$
for $R\neq 0$.
This can be seen from \eq{phi_*}. The same holds true for background
field flows of gauge theories in axial gauges \cite{Litim:2002ce}. For
gauge theories within background field dependent gauges, the situation is
even more involved, since the cut-off term is not the only source for a
non-vanishing right-hand side of \eq{phi_*}. From now on we discuss
the consequences of
\begin{mathletters}\label{keyapprox}
\beq
\Gamma^{(2)}_k[\phi]\stackrel{!}{=}\Gamma^{(2,0)}_k[\phi,\phi] 
\eeq
which is equivalent to the requirement
\beq
2 \Gamma_k^{(1,1)}[\phi,\phi]+\Gamma_k^{(0,2)}[\phi,\phi]
\stackrel{!}{=}0.  
\eeq
\end{mathletters}%
It has to be seen, what results can be obtained within such an
approximation and where it approaches its limits. We emphasise that
\eq{keyapprox} is trivially satisfied without background fields.
Furthermore, in scalar theories and gauge theories within axial gauges
\cite{Litim:2002ce}, the background field only enters the regulator
term. Hence, \eq{phi_*} vanishes in the infra-red limit which
guarantees that \eq{keyapprox} is automatically satisfied in the same
limit. Applications of the flow \eq{flowgeneral} to gauge theories or
gravity in the approximation \eq{keyapprox} and neglecting flow terms
$\sim \dot x$ have been studied in
\cite{Reuter:1994kw,Reuter:1997gx,Reuter:1994sg,ScalarQED-N,QuantumGravity},
while the inclusion of flow terms $\sim \dot x$ has been considered in
\cite{Gies:2002af}.
\bigstep

%********|*********|*********|*********|*********|*********|*********|****
\noindent 
{\bf 3. Exact proper time flows}\label{EPTF}\step
%********|*********|*********|*********|*********|*********|*********|****

Next, we discuss the link of \eq{flowgeneral} to the proper time
renormalisation group (PTRG). It was shown in
\cite{Litim:2001ky,Litim:2002xm}, that PTRG flows are not exact.  For
the derivation we follow the arguments detailed there: a
general PTRG flow at an effective cut-off scale $k^2$ can be expanded
in a set of flows
\begin{eqnarray}
\partial_t\Gamma_{k}[\phi]=\Tr\, \left(
{k^2\over k^2+x/m}\right)^m. 
\label{genPT}\end{eqnarray}
Here, $x=\Gamma_{k}^{(2)}[\phi]$ in the spirit of \eq{defofx}. The
parameter $m$ describes different implementations of the proper-time
regularisation. A general PTRG flow can be seen as a linear
combinations of flows with different $m$. Another difference to
\eq{defofx} stems from the fact, that the proper time flow is derived
without relying on background fields. In \cite{Litim:2002xm} we
showed, that the proper time flows \eq{genPT} represent approximations
to exact background field flows and devised an exact generalised
proper time flow. \Eq{genPT} lacks additional terms
proportional to $t$ derivatives of $\Gamma_k^{(2)}$. Moreover it 
relies on the approximation \eq{keyapprox}.  Now we present the
corresponding $R$ with the leading term as in \eq{genPT}. We choose a
regulator $R$ with $r[x]$ given by
\begin{eqnarray}\label{choice} 
r_m[x]= 
\exp\left(\s01m\left(\s0{mk^2}{x}\right)^m\,{}_2 F_1[m,m;m+1;-\s0{m\, 
k^2}{x}]\right)-1\,. 
\end{eqnarray}
The function $r_m$ is a solution of the differential equation 
\begin{eqnarray}\label{diffdef}\nonumber
\0{x\,r'}{1+r}=-\left(1+\0{x}{m\, k^2}\right)^{-m}\,.
\end{eqnarray}
This equation comes from matching the ERG flow -by neglecting the
terms proportional to $\partial_t\Gamma_k^{(2)}$- to the proper time
flow \eq{genPT}. The flow equation \eq{flowgeneral} with the regulator
function $r$ as in \eq{choice} takes the form
\begin{eqnarray}\label{flowm}
\partial_t \Gamma_k = 
\Tr\, \left({k^2 \over k^2+x/m}\right)^m 
+
{1\over 2} \Tr\left[\, {r_m \over x(1+r_m)}
-\left({k^2 \over k^2+x/m}\right)^m {1\over x}\right]
\,  \dot x\,. 
\end{eqnarray}
\Eq{flowm} reduces to \eq{genPT} if the term proportional to $\dot x$
on the right hand side is dropped. An interesting case is provided by
the limit $m\to \infty$, where $\left(1 +x/(m\,k^2)\right)^{-m}\to
\exp(-x/k^2)$. In this limit $r[x]$ takes the form
\begin{eqnarray}\label{rinfty}
r_\infty[x]= \exp\left(-{\rm Ei}[-\s0{x}{k^2}]\right)-1.
\end{eqnarray}
The flow for the regulator $r_\infty$ reads   
\begin{eqnarray}
\partial_t \Gamma_k = \Tr\, 
\exp\left(-\s0{x}{ k^2}\right)
+\s012\Tr\left[1-\exp\left(-\s0{x}{ k^2}\right)-\exp\left({\rm
Ei}[-\s0{x}{k^2}]\right)\right] \s01x\dot x. 
\label{flowinfty}\end{eqnarray} 
\Eq{flowinfty} boils down to the $m\to\infty$ limit of the proper time flow
\eq{genPT}, when neglecting $\dot x= \partial_t \Gamma_k^{(2)}$. \step

To conclude, apart from terms proportional to $\dot
x=\partial_t\Gamma_k^{(2)}$ the ERG flows \eq{flowm} resemble the
proper time flows \eq{genPT}.  Therefore, a general proper time flow
like \eq{genPT} is the approximation to a background field ERG flow,
where $(i)$ terms proportional to $\dot x$ are neglected and 
$(ii)$ implicitly the difference between fluctuation field and background field
has been neglected, resorting to \eq{keyapprox}. \step

This suggests to compare results of the (non-exact) PTRG to results
from flows with the regulator argument $\Gamma_k^{(2)}$. Here, one can
use optimisation criteria for this set of regulators
\cite{Litim:2000ci}. Note that the choice \eq{choice} was only one
particular example of the general set of $\Gamma^{(2)}_k$-dependent
regulators. Applications of the proper-time RG are based on flows
\eq{genPT}, and approximations thereof. It would be interesting to see
how these results are affected by the additional flow terms as given
in \eq{flowm}.\step

The above analysis is also linked to a further observation made in
\cite{Litim:2002xm}. The flows \eq{genPT} and \eq{flowm} can be
matched to generalised Callan-Symanzik flows. The difference has two
sources: First, the flows deviate by terms proportional to
$\partial_t\Gamma_k^{(2)}$ and higher scale derivatives of $\Gamma_k$
itself. Second, the Callan-Symanzik flow is defined without background
fields and the argument $x=\Gamma_k^{(2)}$ is just the propagator of
the full field.  This has to be compared to \eq{flowm}, which is not a
closed equation, and to \eq{genPT}, which is not exact. This leaves us
with a representation of the missing terms made by the approximation
\eq{keyapprox} without resorting to background fields at all. \bigstep

%********|*********|*********|*********|*********|*********|*********|****
\noindent 
{\bf 4. Implicit scale dependences}\step
%********|*********|*********|*********|*********|*********|*********|****

Up to now, we have discussed the implications of using
a background field dependent regularisation. In general, we are led 
to flows that contain terms proportional to
$\partial_t\Gamma^{(2)}$. A priori, it is difficult to estimate the numerical
importance of those additional flow terms. Moreover, 
the validity of the approximation \eq{keyapprox}
cannot be taken for granted.  Hence, it would be most useful to gain
some intuition for these matters, in particular in view of future
numerical applications.  Here, we study the relevance of
$\partial_t\Gamma^{(2)}$ terms for $O(N)$ symmetric scalar theories at
criticality in three dimensions, using ERG flows with background
fields in different approximations. 
In the following we discuss flows for regulators of the form
\begin{eqnarray}\label{variant} 
R=
(k^2(1-\X)-q^2) \,\theta
(k^2(1-\X)-q^2), 
\end{eqnarray}
where $q^2$ is plain momentum squared and $\X$ encodes some intrinsic
scale or background field dependence. For $\X=0$, it reduces to an
optimised regulator as introduced in \cite{Litim:2000ci}. We consider
the flow equation for the effective potential $U_k$ to leading order
in the derivative expansion.  The flow equation is best written in
dimensionless variables $u=U_k\, k^{-d}$ and $\rho=\s012\phi^a\phi_a\,
k^{2-d}$.  For the choice \eq{variant} of the regulator, the momentum
integration can be performed analytically and leads to
\bea
\nonumber 
\partial_t u+d u -(d-2) \rho u' &=&
(N-1)\, 
\0{(1-\X_1)^{d/2}(1-\X_1-\s012\partial_t \X_1)}{1-\X_1+u'}\
\theta(1-\X_1)\\
& &  
+\,\0{(1-\X_2)^{d/2}(1-\X_2-\s012\partial_t \X_2)}{1-\X_2 +u'+2\rho u''}\
\theta(1-\X_2)\, 
\label{variantflow}
\eea
where $\X_1, \X_2$ are Goldstone and radial parts of the matrix $\X$
in field space. For simplicity, we have rescaled the irrelevant factor
$4v_d/d$ on the right-hand side of \eq{variantflow} into the potential
and the fields.\step

Since $\X$ is at our disposal it can be used to enhance the stability
of the flow. The standard case $X=0$ (plain momentum regularisation)
has been discussed in \cite{Litim:2002cf}. We discuss the following
natural choices for $\X$:\step 

\begin{itemize}
\item[$A$\ \ ] One can consider a regularisation which naturally
  includes an expansion about the minimum of the effective potential.
  We choose $q^2+\X=\Gamma_k^{(2,0)}[\phi_0,\phi_0](q^2)$.
  Additionally, we take $\phi_0=0$, which defines the $k$-dependent
  mass $m^2_k$ at vanishing field. We have
\begin{eqnarray}
\X_1=\X_2=m_k^2/k^2.
\label{choiceA} 
\end{eqnarray}
The flow \eq{variantflow} with the choice \eq{choiceA} is denoted as
flow $A$. Here, the regularisation introduces additional
scale-dependent contributions to the flow, triggered by the running of
the mass term at vanishing field. In consequence, no explicit
background field dependence is introduced due to the regulator.

\item[$B$\ \ ] A second choice for $\X$ is given by
  $q^2+\X=\Gamma_k^{(2,0)}[\phi,\phi]$, the full field-dependent
  propagator. This entails that our flow is essentially diagonal, the
  operator trace only involves the full propagator and its scale
  derivative. However, this forces us to concentrate on the flow of
  $\Gamma_k[\phi]$ as defined in the previous section in \eq{onefield}
  within the approximation \eq{keyapprox}. We have
\begin{eqnarray}
\X_1=u'(\rho), &\qquad & \X_2=u'(\rho)+2\rho u''(\rho) \,.
\label{choiceB} 
\end{eqnarray}
The flow \eq{variantflow} with the choice \eq{choiceB} is denoted as
flow $B$. Notice that the choice \eq{choiceB} corresponds to a cutoff
in momenta and a cutoff in field amplitude, because the flow is
suppressed once $\Gamma^{(2)}_k>k^2$. In contrast to flow $A$, the
regularisation of the flow $B$ depends explicitly on the background
field.

\item[$C$\ \ ] Finally, we also compare the flows $A$ and $B$ to the
  exact proper time flow \eq{flowinfty} for $m=\infty$.  The operator
  $\X$ is given by \eq{choiceB}, and the flow is referred to as the
  flow $C$. It differs from flow $B$ only in the choice for the
  regularisation.
\end{itemize}

The structure of the flow \eq{variantflow} depends significantly on
the choices $X=0$, \eq{choiceA} or \eq{choiceB}. For $X=0$, the flow
has a potential pole when $u'$ or $u'+2\rho u''$ approach $-1$. This is
a direct consequence of the IR regularisation. For \eq{choiceA}, the
pole is shifted away because the choice for $X$ corresponds precisely
to the most negative value for $u'$ or $u'+2\rho u''$, respectively.
This is even more pronounced for the field-dependent choice
\eq{choiceB}, because the denominator in \eq{variantflow} become
trivial and field-independent. Still, the additional flow terms on the
right-hand side of \eq{variantflow} imply potential poles,
e.g.~\cite{Litim:2002cf}.
\step

Next, we study the impact of the implicit scale dependence
on the flow. This is done by partially considering the additional
terms coming from the scale derivatives of $\X$ in \eq{variantflow}.
The full scale derivative is proportional to
\begin{eqnarray}\label{scalederiv} 
k^{-2}\partial_t (k^2 \X)=2\X+\partial_t \X\,. 
\end{eqnarray} 
The first term on the right-hand side displays the explicit scale
dependence of $\sim k^2$, whereas the second term contains the
implicit scale dependence of $\X$ bound to vanish at a fixed point.
Our first approximation is based on dropping the entire scale
derivative of $k^2\X$ by setting
\begin{eqnarray}
k^{-2}\partial_t (k^2 \X)=0. 
\label{ell-mod2}\end{eqnarray} 
This is referred to as flows $A'$, $B'$ and $C'$, the prime indicating
the additional approximation \eq{ell-mod2} on $A$, $B$ and $C$. 
In particular, the flow
$C'$ reads
\begin{eqnarray}\label{potm=infty} 
\partial_t u+d u -(d-2) \rho u' =
(N-1)\, \exp\left(-u'\right) +\, \exp\left(-u'-2\rho u''\right)\,, 
\end{eqnarray}
where we have absorbed an irrelevant numerical factor $2v_d$ into the
fields. This flow has been studied numerically in
\cite{Mazza:2001bp,Litim:2001hk}.\footnote{3d critical exponents have
  also been computed from \eq{genPT} for other values of $m$,
  e.g.~\cite{PTRG-Applications,Bonanno:2001yp,Litim:2001hk}.}  The
second kind of approximation consists in neglecting the intrinsic
dependence on $\partial_t\X$, defined by
\begin{eqnarray}
k^{-2}\partial_t (k^2 \X)=2\X\,. 
\label{ell-mod3}\end{eqnarray}
This is referred to as flows $A''$, $B''$ and $C''$, the double-prime
indicating the additional approximation \eq{ell-mod3}. \step

On a fixed point, we have $\partial_t X=0$, because $X$ is a
dimensionless function. Using \eq{scalederiv} and \eq{ell-mod3}, we
conclude that the fixed point solutions of the unprimed and the
double-primed flows are identical. More generally, the fixed point
solutions of any two flows are identical, if the flows differ only by
a dimensionless flow term, which vanishes on a fixed point. For the
$A$-flows, the function $X$ is field-independent. From the explicit
form of the flows $A$ and $A'$, we conclude that the non-universal
fixed point solutions $\partial_t u=0$ are related by a simple
rescaling of the fields and the effective potential.  An analogous
statement cannot be made about universal critical exponents.  Indeed,
critical exponents describe the approach towards the critical point.
In the vicinity of the fixed point, the RG trajectories are strongly
sensitive to the functions \eq{scalederiv}, \eq{ell-mod2} or
\eq{ell-mod3}.  Despite the close similarity or even equality of the
fixed point solutions, in general their critical exponents are all
different.  \bigstep

%********|*********|*********|*********|*********|*********|*********|****
\noindent 
{\bf 5. Results}\step
%********|*********|*********|*********|*********|*********|*********|****

In Tab.~1, we have computed the critical exponent $\nu$ for the flows
$A$, $A'$, $A''$, $B'$, $B''$ and $C'$ for all $N$ between $-2$ and
$\infty$.  Results have been given for the physically most interesting
cases $N=0,1,2,3$ and $4$. The critical exponent $\nu$ describes the
approach towards the critical point.  In the vicinity of the fixed
point, the RG trajectories are strongly sensitive to the explicit form
of the flow, which is quite different for the different cases
introduced above. For $N=-2$, the critical exponent is known to be
$\nu=\s012$. The universal large-$N$ result reads $\nu=1$. All flows
in all approximations reproduce the known result at $N=-2$.  Except
for the flows $A'$ and $A''$, the same holds also true for the
large-$N$ limit. All results given in Tab.~1 are within $20\%$ of the
physical values, and some results are significantly closer. The
results also depend on the choice of $R$. Within the present
truncation the variation with $R$ is of a similar order of magnitude
\cite{Litim:2002cf}. \step

%********|*********|*********|*********|*********|*********|*********|****
%********|*********| Tab1    |*********|*********|*********|*********|****
%********|*********|*********|*********|*********|*********|*********|****
\begin{center}
\begin{tabular}{c||c|c|c||c|c||c}
$\quad N\quad$&
$\quad\quad A\quad\quad$&
$\quad\quad A'\quad\quad $&
$\quad\quad A''\quad\quad $&
$\quad\quad B'\quad\quad $&
$\quad\quad B''\quad \quad$&
$\quad\quad C'\quad\quad $
\\[.5ex] \hline%\\[-1.5ex]
$-2$&
$\s012$&
$\s012$&
$\s012$&
$\s012$&
$\s012$&
$\s012$\\[.5ex]
0&
.592&
.583&
.649&
.565&
.574&
.582\\[.5ex]
1&
.650&
.626&
.738&
.587&
.607&
.626\\[.5ex]
2&
.708&
.666&
.825&
.605&
.636&
.669\\[.5ex]
3&
.761&
.699&
.901&
.620&
.663&
.710\\[.5ex]
4&
.804&
.725&
.962&
.631&
.688&
.749\\[.5ex]
$\infty$&
1&
.828&
1.21&
1&
1&
1
\end{tabular}
\end{center}
\begin{center}
\begin{minipage}{.95\hsize}
  \vskip.3cm {\small {\bf Table 1:} Critical exponents $\nu$ to
    leading order in the derivative expansion from various flows. \step}
\end{minipage}
\end{center}
%********|*********|*********|*********|*********|*********|*********|****

On a technical level, the integration of the flows $A$, $B$ and $C$
is significantly more difficult than the standard flow $X=0$ or the
primed flows, because they involve higher order flow terms on their
right-hand sides. For the flow $A$, this dependence is simple enough
to be diagonalised explicitly \cite{Litim:2002cf}. For the flow $B$,
the study is simplified by noticing that the fixed point solution of
$B$ and $B''$ are identical.  Once the fixed point solution is found,
the critical exponents can be computed directly. For the computations,
we make use of a polynomial approximation, either about vanishing
field or the minimum of the potential. The critical exponents showed
good convergence except for the flow $B$, where no definite result has
been found within the polynomial expansion.\step

Consider the cases $A$, $A'$ and $A''$.  By construction, all
$A$-flows automatically fulfil the requirement \eq{keyapprox}.
Comparing the results for $A$ and $A'$, we conclude that the intrinsic
scaling contributes between $2-10\%$ to the value for $\nu$.  Notice
that the values in the second row are slightly closer to the physical
values \cite{Litim:2001dt}, despite the fact that the underlying
approximation is less accurate than the one leading to the flow $A$.
In the flow $A''$, we have neglected only the term in \eq{scalederiv}
proportional to $\partial_t(m^2_k/k^2)$. In comparison to $A$ and
$A'$, we notice a strong increase for the critical exponent about
$10-20\%$. Conversely, retaining only the term $\propto \partial_t
(m^2_k/k^2)$ of \eq{scalederiv} in the flow would have lead to a
decrease of $\nu$ by roughly the same amount (data not displayed).
Hence, we found that {\it both} terms of \eq{scalederiv} contribute
strongly and with opposite sign to universal quantities. We also
notice that the limit $N=-2$ more stable than the large-$N$ limit,
which is not reproduced by $A'$ and $A''$. This shows that maintaining
\eq{keyapprox} and the correct large-$N$ limit is not compatible with
the approximations \eq{ell-mod2} or \eq{ell-mod3}.  \step

Next we consider the flows $B'$ and $B''$. By construction, all
$B$-flows are subject to the approximation \eq{keyapprox}. The
$B$-flows are sensitive to a non-trivial background field dependence
of the regulator. The flows $B'$ and $B''$ reproduce the correct
limits for $N=-2$ and $\infty$. This shows that the approximation
\eq{keyapprox} is compatible with \eq{ell-mod2} or \eq{ell-mod3} for
background field dependent regulators, even in the large-$N$ limit.
Numerically, the results for $B'$ and $B''$ are below the physical
values. Also, the values for $B'$ are smaller than those for $B''$.
This is qualitatively the same as for the $A$-flows.\step

Structurally, the flow $C'$ differs from $B'$ only in the choice for
the regulator, which is given by \eq{R-background} with \eq{rinfty} in
the former, and by \eq{variant} with \eq{choiceB} in the latter. The
flow $C'$ differs from $A'$ in the additional background field
dependence. This affects the numerical values for the critical
exponents. This does not affect the limits $N=-2$ or $N=\infty$, in
agreement with the reasoning given above for the $B$-flows. The $C'$
results agree with those given in \cite{Mazza:2001bp,Litim:2001hk}. It
is very intriguing that the values obtained for $A'$ and $C'$ are
both very close to the physical values, and very close to each other,
ranging within $.2-3\%$ for $N\in[0,4]$. In the light of our
discussion above, we expect that the additional terms found in
\eq{flowm} should modify the results for the critical indices.
\bigstep

%********|*********|*********|*********|*********|*********|*********|****
\noindent 
{\bf 6. Conclusions}\step 
%********|*********|*********|*********|*********|*********|*********|****

We have studied the implications of background field dependent
flows. Conceptually, the r$\hat{\rm o}$le of $\Gamma^{(2,0)}$ has been
threefold. First of all, its appearance in the regulator has lead to
an important simplification. The background field flow becomes
essentially diagonal leading to a natural expansion in the
eigenfunctions of $\Gamma^{(2,0)}$. Secondly, the deviation from this
property is proportional to $\partial_t \Gamma^{(2,0)}$ leading to
additional flow terms. Their numerical treatment was more difficult
 than the standard flow. Thirdly, the regulator cuts off
both large momentum and large field amplitude contributions to the
flow. We expect that the background field flows presented here are
useful for examining `local' properties such as critical exponents as
done in the present case. Global flows have to be studied separately
as more care is required concerning the irrelevance of the neglected terms. 
\step 

Numerically, a coherent picture has emerged as well. The implicit
dependences leave the qualitative result unaffected, supporting the
viability of the approximation \eq{keyapprox}. Still, the corrections
are quantitatively large and cannot be neglected for high precision
computations. Best agreement with experimental data is achieved for
the flows $A'$ and $C'$, despite the fact that implicit dependences
are suppressed in both cases. This deserves further study, in
particular in view of the full background field dependence. \step

More work has to be done for a study of non-universal quantities and
for applications to non-Abelian gauge theories. There, most notably,
the anomalous dimension of the gauge field is far from being small,
the latter being a property coming to our help in the present work.
\bigstep

%********|*********|*********|*********|*********|*********|*********|****
\noindent 
{\bf Acknowledgements}\step 
%********|*********|*********|*********|*********|*********|*********|****

JMP thanks CERN for hospitality and financial support.  The work of
DFL has been supported by the European Community through
the Marie-Curie fellowship HPMF-CT-1999-00404.\bigstep

%********|*********|*********|*********|*********|*********|*********|****


\begin{thebibliography}{99}
%********|*********|*********|*********|*********|*********|*********|****

%\cite{Reuter:1994kw}
\bibitem{Reuter:1994kw}
M.~Reuter and C.~Wetterich,
%``Effective average action for gauge theories and exact evolution
%equations,''
Nucl.\ Phys.\  {\bf B417} (1994) 181.
%%CITATION = NUPHA,B417,181;%%

%\cite{Reuter:1997gx}
\bibitem{Reuter:1997gx}
M.~Reuter and C.~Wetterich,
%``Gluon condensation in nonperturbative flow equations,''
Phys.\ Rev.\ D {\bf 56} (1997) 7893
[hep-th/9708051].
%%CITATION = HEP-TH 9708051;%%


%\cite{Reuter:1995tr}
\bibitem{Reuter:1995tr}
M.~Reuter,
%``Effective average action of Chern-Simons field theory,''
Phys.\ Rev.\ D {\bf 53} (1996) 4430
[hep-th/9511128].
%%CITATION = HEP-TH 9511128;%%


%\cite{Reuter:1996be}
\bibitem{Reuter:1996be}
M.~Reuter,
%``Renormalization of the Topological Charge in Yang-Mills Theory,''
Mod.\ Phys.\ Lett.\ A {\bf 12} (1997) 2777
[hep-th/9604124].
%%CITATION = HEP-TH 9604124;%%

\bibitem{Pawlowski:1998ch}
J.~M.~Pawlowski,
%``Exact Flow Equations and the $U(1)$-Problem,''
Phys.\ Rev.\  {\bf D58} (1998) 045011
[hep-th/9605037].
%%CITATION = HEP-TH 9605037;%%

\bibitem{Litim:1998qi}
%\cite{Litim:1998qi}
D.~F.~Litim and J.~M.~Pawlowski,
%``Flow equations for Yang-Mills theories in general axial gauges,''
Phys.\ Lett.\  {\bf B435} (1998) 181
[hep-th/9802064];
%%CITATION = HEP-TH 9802064;%%
%\cite{Litim:1999yc}
%D.~F.~Litim and J.~M.~Pawlowski,
%``On general axial gauges for {QCD},''
Nucl.\ Phys.\ Proc.\ Suppl.\  {\bf 74} (1999) 329
[hep-th/9809023]; 
%%CITATION = HEP-TH 9809023;%%
%``On gauge invariance and Ward identities for the Wilsonian
%renormalisation group,''
Nucl.\ Phys.\ Proc.\ Suppl.\  {\bf 74} (1999) 325
[hep-th/9809020].
%%CITATION = HEP-TH 9809020;%%

%\cite{Falkenberg:1998bg}
\bibitem{Falkenberg:1998bg}
S.~Falkenberg and B.~Geyer,
%``{{\it Effective average action in N = 1 super-Yang-Mills theory,}}''
Phys.\ Rev.\  {\bf D58} (1998) 085004
[hep-th/9802113].
%%CITATION = HEP-TH 9802113;%%



%\cite{Litim:1998nf}
\bibitem{Litim:1998nf}
D.~F.~Litim and J.~M.~Pawlowski,
%``On gauge invariant Wilsonian flows,''
hep-th/9901063.
%%CITATION = HEP-TH 9901063;%%


%\cite{Freire:1996db}
\bibitem{Freire:1996db}
F.~Freire and C.~Wetterich,
%``The Ward Identity from the Background Field Dependence of the Effective Action,''
Phys.\ Lett.\ B {\bf 380} (1996) 337
[hep-th/9601081].
%%CITATION = HEP-TH 9601081;%%

%\cite{Freire:2000bq}
\bibitem{Freire:2000bq}
F.~Freire, D.~F.~Litim and J.~M.~Pawlowski,
%``Gauge invariance and background field formalism in the exact  renormalisation group,''
Phys.\ Lett.\ B {\bf 495} (2000) 256
[hep-th/0009110];
%%CITATION = HEP-TH 0009110;%%
%\cite{Freire:2001mn}
%\bibitem{Freire:2001mn}
%F.~Freire, D.~F.~Litim and J.~M.~Pawlowski,
%``Gauge invariance, background fields and modified Ward identities,''
Int.\ J.\ Mod.\ Phys.\ A {\bf 16} (2001) 2035
[hep-th/0101108].
%%CITATION = HEP-TH 0101108;%%



%\cite{Pawlowski:2001df}
\bibitem{Pawlowski:2001df}
J.~M.~Pawlowski,
%``On Wilsonian flows in gauge theories,''
Int.\ J.\ Mod.\ Phys.\ A {\bf 16} (2001) 2105; in preparation. 
%%CITATION = IMPAE,A16,2105;%%



%\cite{Bonini:2001xm}
\bibitem{Bonini:2001xm}
M.~Bonini and E.~Tricarico,
%``{{\it Background field method in the Wilson formulation,}}''
Nucl.\ Phys.\ B {\bf 606} (2001) 231
[hep-th/0104255].
%%CITATION = HEP-TH 0104255;%%






%\cite{Gies:2002af}
\bibitem{Gies:2002af}
H.~Gies,
%``Running coupling in Yang-Mills theory: A flow equation study,''
Phys.\ Rev.\ D {\bf 66} (2002) 025006
[hep-th/0202207].
%%CITATION = HEP-TH 0202207;%%



%\cite{Litim:2002ce}
\bibitem{Litim:2002ce}
D.~F.~Litim and J.~M.~Pawlowski,
%``Renormalisation group flows for gauge theories in axial gauges,''
hep-th/0203005.
%%CITATION = HEP-TH 0203005;%%




%\cite{Reuter:1994sg}
\bibitem{Reuter:1994sg}
M.~Reuter and C.~Wetterich,
%``Exact evolution equation for scalar electrodynamics,''
Nucl.\ Phys.\ B {\bf 427} (1994) 291.
%%CITATION = NUPHA,B427,291;%%



%\cite{ScalarQED-N}
\bibitem{ScalarQED-N}
B.~Bergerhoff, D.~F.~Litim, S.~Lola and C.~Wetterich,
Int.\ J.\ Mod.\ Phys.\  {\bf A11} (1996) 4273;\\
%%CITATION = COND-MAT 9502039;%%
%\bibitem{ScalarQED}
B.~Bergerhoff, F.~Freire, D.~F.~Litim, S.~Lola and C.~Wetterich,
Phys.\ Rev.\  {\bf B53} (1996) 5734;\\
%%CITATION = HEP-PH 9503334;%%
%\cite{Freire:2001sx}
%\bibitem{Freire:2001sx}
F.~Freire and D.~F.~Litim, 
Phys.\ Rev.\ D {\bf 64} (2001) 045014 [hep-ph/0002153].
%%CITATION = HEP-PH 0002153;%%





%\cite{QuantumGravity}
\bibitem{QuantumGravity}
%\cite{Reuter:1996cp}
%\bibitem{Reuter:1996cp}
M.~Reuter,
%``Nonperturbative Evolution Equation for Quantum Gravity,''
Phys.\ Rev.\ D {\bf 57} (1998) 971
[hep-th/9605030];\\
%%CITATION = HEP-TH 9605030;%%%\cite{Lauscher:2001ya}
%\bibitem{Lauscher:2001ya}
O.~Lauscher and M.~Reuter,
%``Ultraviolet fixed point and generalized flow equation of quantum  gravity,''
Phys.\ Rev.\ D {\bf 65} (2002) 025013
[hep-th/0108040];\\
%%CITATION = HEP-TH 0108040;%%
%\cite{Lauscher:2001rz}
%\bibitem{Lauscher:2001rz}
%O.~Lauscher and M.~Reuter,
%``Is quantum Einstein gravity nonperturbatively renormalizable?,''
%Class.\ Quant.\ Grav.\  {\bf 19} (2002) 483
%[hep-th/0110021];
%%CITATION = HEP-TH 0110021;%%
%\cite{Reuter:2001ag}
%\bibitem{Reuter:2001ag}
M.~Reuter and F.~Saueressig,
%``Renormalization group flow of quantum gravity in the Einstein-Hilbert  truncation,''
Phys.\ Rev.\ D {\bf 65} (2002) 065016
[hep-th/0110054].
%%CITATION = HEP-TH 0110054;%%


%\cite{Liao:1997nm}
\bibitem{Liao:1997nm}
S.~B.~Liao,
%``Operator Cutoff Regularization and Renormalization Group in Yang-Mills Theory,''
Phys.\ Rev.\  {\bf D56} (1997) 5008
[hep-th/9511046];
%%CITATION = HEP-TH 9511046;%%
%\cite{Liao:1996fp}
%\bibitem{Liao:1996fp}
%S.~B.~Liao,
%``On connection between momentum cutoff and the proper time regularizations,''
Phys.\ Rev.\  {\bf D53} (1996) 2020
[hep-th/9501124].
%%CITATION = HEP-TH 9501124;%%


%\cite{Litim:2001hk}
\bibitem{Litim:2001hk}
D.~F.~Litim and J.~M.~Pawlowski,
%``Predictive power of renormalisation group flows: A comparison,''
Phys.\ Lett.\ B {\bf 516} (2001) 197
[hep-th/0107020].
%%CITATION = HEP-TH 0107020;%%



%\cite{Litim:2001ky}
\bibitem{Litim:2001ky}
D.~F.~Litim and J.~M.~Pawlowski,
%``Perturbation theory and renormalisation group equations,''
Phys.\ Rev.\ D {\bf 65} (2002) 081701
[hep-th/0111191].
%%CITATION = HEP-TH 0111191;%%


%\cite{Litim:2002xm}
\bibitem{Litim:2002xm}
D.~F.~Litim and J.~M.~Pawlowski,
%``Completeness and consistency of renormalisation group flows,''
Phys.\ Rev.\ D {\bf 66} (2002) 025030
[hep-th/0202188].
%%CITATION = HEP-TH 0202188;%%

%\cite{Litim:2000ci}
\bibitem{Litim:2000ci}
D.~F.~Litim,
%``Optimisation of the exact renormalisation group,''
Phys.\ Lett.\  {\bf B486} (2000) 92
[hep-th/0005245]; 
%%CITATION = HEP-TH 0005245;%%
%``Optimised renormalisation group flows,''
Phys.\ Rev.\ D {\bf 64} (2001) 105007
[hep-th/0103195]; 
%%CITATION = HEP-TH 0103195;%%
%``Mind the gap,''
Int.\ J.\ Mod.\ Phys.\ A {\bf 16} (2001) 2081
[hep-th/0104221]; 
%%CITATION = HEP-TH 0104221;%%
%``Scheme independence at first order phase 
%transitions and the  renormalisation group,''
Phys.\ Lett.\  {\bf B393} (1997) 103 [hep-th/9609040].
%%CITATION = HEP-TH 9609040;%%


%\cite{Litim:2002cf}
\bibitem{Litim:2002cf}
D.~F.~Litim,
%``Critical exponents from optimised renormalisation group flows,''
Nucl.\ Phys.\ B {\bf 631} (2002) 128
[hep-th/0203006].
%%CITATION = HEP-TH 0203006;%%

%\cite{Mazza:2001bp}
\bibitem{Mazza:2001bp}
M.~Mazza and D.~Zappal\`a,
%``Proper time regulator and renormalization group flow,''
Phys.\ Rev.\ D {\bf 64} (2001) 105013
[hep-th/0106230].
%%CITATION = HEP-TH 0106230;%%

\bibitem{PTRG-Applications}
O.~Bohr, B.~J.~Sch\"afer and J.~Wambach,
%``Renormalization group flow equations and the phase transition in O(N)
%models,''
Int.\ J.\ Mod.\ Phys.\ A {\bf 16} (2001) 3823
[hep-ph/0007098].
%%CITATION = HEP-PH 0007098;%%

%\cite{Bonanno:2001yp}
\bibitem{Bonanno:2001yp}
A.\,Bonanno and D.\,Zappal\`a,
%``Towards an accurate determination of the critical exponents with the
%renormalization group flow equations,''
Phys.~Lett.~{\bf B504} (2001) 181
[hep-th/0010095].
%%CITATION = HEP-TH 0010095;%%

%\cite{Litim:2001dt}
\bibitem{Litim:2001dt}
D.~F.~Litim,
%``Derivative expansion and renormalisation group flows,''
JHEP {\bf 0111} (2001) 059
[hep-th/0111159].
%%CITATION = HEP-TH 0111159;%%


\end{thebibliography}
\end{document}